\documentstyle [12pt]  {article}
 \evensidemargin\oddsidemargin

\begin{document}
 \title{{ Nuclear Decay Parameter  Oscillations as Possible \\
Signal of Quantum Nonlinearity \\ }}
\author {{ S.N.Mayburov} \\
 Lebedev Inst. of Physics\\
   Leninsky Prospect 53,  Moscow, Russia, RU-117924\\
  e-mail: mayburov@sci.lebedev.ru \\}
 \date { }
 \maketitle



%
\begin{abstract}
 Several experimental groups reported the evidence of multiple periodic modulations of
nuclear decay constants which amplitudes are of the order
$10^{-3}$ and  periods of one year, 24 hours or about one month.
We argue that such deviations from radioactive decay law can be
described in nonlinear quantum mechanics framework, in which decay
process obeys to nonlinear Shroedinger equation with Doebner-Goldin terms. 
Proposed corrections to Hamiltonian of quantum
system interaction with gravitation field 
 correspond to some emergent gravity theories, in
particular, bilocal gravity model. 
 Decay parameter variations under influence of Sun gravity,
  calculated in our model,  agree well with experimental results for $\alpha$-decay
life-time oscillations of Polonium isotopes.


\end{abstract}

\section{Introduction}
\label{intro}

Natural radioactivity law is one of most fundamental laws
of modern physics, in accordance with it,
  nuclear decay parameters
 are time-invariant and practically independent of environment \cite {Mar}.
  However, some  recent experiments have reported the evidence of periodic modulations of
nuclear $\alpha-$ and $\beta-$decay parameters of the order of
$10^{-3}$ and with typical periods of one year, one day or about
one month [2-8].
 Possible mechanism of such decay parameter
oscillations is still unclear,
   explanations proposed until now
don't look convincing \cite {Fis}. Therefore, it's sensible to
start the effect analysis from reconsideration of nuclear decay
fundamentals.
 In this paper,
these oscillation effects studied in the framework of
quantum-mechanical theory of unstable system decay. It follows
 from our analysis that standard quantum formalism can't explain
 the observed parameter oscillations.
 However, it will be  argued  that nonlinear
modifications of quantum mechanics, extensively studied in last
years [9-12], in particular, nonlinear interactions of quantum systems with gravity,
presumably can describe the decay parameter variations with the
similar features.


 First results, indicating the essential deviations from  exponential $\beta$-decay rate
  dependence, were obtained during the precise measurement of
$^{32}$Si isotope life-time  \cite {Alb}. Sinusoidal annual
oscillations with the amplitude  $6*10^{-4}$ relative to total
decay rate and maxima located at the end of February were
registrated during 5 years of measurements. Since then,
  the annual oscillations of
 $\beta$-decay rate for different heavy nuclei from Ba to Ra
 were reported,
for most of them the  oscillation amplitude is of the order
$5*10^{-4}$ with its maximum on the average at mid-February \cite
{Fis}. Annual oscillations of $^{238}$Pu $\alpha$-decay rate with
the amplitude of the order $10^{-3}$ also were reported \cite
{El}.
  Life-time of short-living $\alpha$-decayed isotopes
$^{214}$Po, $^{213}$Po  was measured directly, the annual and daily
oscillations with amplitude of the order $9*10^{-4}$, with annual
maxima at mid-March and daily maxima around 6 a.m. were found
during five years of measurements \cite {Al}. Some other effects
related to nucleus decay oscillations also described in the
literature. Annual and daily oscillation periods were found in
the studies of $^{239}$Pu $\alpha$-decay statistics \cite
{Sh,Sh2}. Small oscillations of decay electron energy spectra
with period 6 months were found in Tritium $\beta$-decay \cite
{Lob}.
 Some other $\beta$-decay
experiments  exclude any decay constant modulations as large as
reported ones \cite {Kos,Bel}.

 Until now theoretical discussion of these
results had quite restricted character. In particular,
oscillations of $\beta$-decay rate was  hypothized as  anomalous
interaction of Sun neutrino flux with nuclei or seasonal
variations of fundamental constants \cite {Fis}.
 Yet, neither of these
hypothesis can explain $\alpha$-decay parameter oscillations of
the same order, because nucleus $\alpha$-decay should be
insensitive to neutrino flux or other electro-weak processes.
Really, $\alpha-$ and $\beta$-decays stipulated by nucleon strong
and weak interactions correspondingly. Therefore, observations of
parameter oscillations for both decay modes supposes that some
universal mechanism independent of particular type of  nuclear
interactions  induces the decay parameter oscillations.

Nowadays, the most universal microscopic theory is quantum
mechanics (QM), so it's worth to start the study of these effects
from reminding  quantum-mechanical description of nucleus decay
\cite {Lan}. In its framework, radioactive nucleus treated as the
metastable quantum state,  evolution of such states was the
subject of many investigations and its principal features are now
well understood \cite {Fon}.  However, due to serious mathematical
difficulties,  precise  calculations of decay processes aren't
possible, and due to it, the simple semi-qualitative models are
used. In particular, some decays modes of heavy nuclei can be
effectively described  as the quantum tunneling of decay products
through the potential barrier constituted by nuclear shell and
nucleus Coulomb field. The notorious example, is Gamow theory of
$\alpha$-decay which describes successfully its main features, as
well as some other decay modes \cite {Gam,New}.
 However, in its standard formulation,
Gamow theory excludes any significant variations of decay
parameters under influence of Sun gravity and similar factors. In
this paper, it's argued that such influence can  appear, if one
applies for $\alpha$-decay description  the nonlinear
modification of standard QM, which developed extensively in the
last years \cite {Bl,Wn}. In particular, we shall use
Doebner-Goldin nonlinear QM model for the description of gravity
 influence on nucleus decay parameters
\cite {DG,DG2}. Basing on its ansatz, Gamow $\alpha$-decay theory with
nonlinear Hamiltonian corrections will be constructed, its model
calculations compared with experimental results for $^{214}$Po
$\alpha$-decay life-time variations \cite {Al}.  In this framework, 
observed influence of Sun gravity on nuclear decay parameters
corresponds to  some emergent gravity theories \cite
{Diaz}.

     \section {Nonlinear QM model}
Interest to nonlinear  evolution equations can be dated back to
the early days of quantum physics, but at that time they appeared
in effective theories describing the collective effects \cite
{Lan}. Now it's acknowledged  that  nonlinear corrections to
standard QM Hamiltonian can exist also at fundamental level \cite
{Jauch,Jor}. Significant progress in the
 studies of such nonlinear QM formalism was achieved in the 80s,
marked by the seminal papers of Bialynicki-Birula and Mycielcki
(BM), and Weinberg \cite {Bl,Wn}. Since then,  many variants of
nonlinear QM formalism were considered in the literature (see \cite {DG2}
and refs. therein). Some experimental tests of QM nonlinearity
were performed, but they didn't have universal character, rather
they tested Weinberg and BM models only \cite {Jor}.

 Currently,
there are two different approaches to the nature of QM
nonlinearity. The main and historically first one supposes that
dynamical nonlinearity is universal and generic property of
quantum systems \cite {Bl,Wn}. In particular, it means that
nonlinear evolution terms can influence their free motion,
inducing soliton-like corrections to standard QM wave packet
evolution \cite {Bl}. Alternative concept of QM nonlinearity
which can be called interactive, was proposed by Kibble, it
postulates that free system evolution should be principally
linear, so that nonlinear dynamics related exclusively to the
system interactions with the fields or field self-interaction
\cite {Kib}. Until now, detailed calculations of such nonlinear
effects were performed only for hard processes of particle
production in the strong fields \cite {Kib2,Elz}, this formalism
can't be applied directly to nonrelativistic processes, like
nuclear decay. Due to it, to describe the interaction of
metastable state with external field, we'll start from
consideration of universal nonlinear models and develop their
modification, which can incorporate the nonlinear particle-field
and nucleus-field interactions at low energies.

In  the universal approach to QM nonlinearity,
it supposed that the particle evolution described by nonlinear
Schroedinger equation of the form \cite {Lan}
\begin {equation}
   i\hbar \partial_t \psi=   -\frac{\hbar^2}{2m}\bigtriangledown^2\psi +V(\vec{r},t)\psi
    + F(\psi,\bar\psi)\psi \label {AXA}
\end {equation}
where $m$ is particle mass, $V$ is external potential, $F$ is
arbitrary functional of system state.
Currently, the most popular and elaborated  nonlinear QM model of
universal type is by Doebner and Goldin (DG) \cite {DG,DG2}, in
its formalism, the simplest variant of nonlinear term 
\begin {equation}
    F= \hbar^2 \lambda (\bigtriangledown^2 + \frac{|\bigtriangledown\psi|^2 }{|\psi|^2} ) \label {BXB}
\end {equation}
where $\lambda$ is imaginary  constant. With the notation
$$
   H_0=-\frac{\hbar^2}{2m}\bigtriangledown^2 +V(\vec{r},t)
$$
we abbreviate (\ref{AXA}) to
$i\hbar{\partial_t\psi}=H_0\psi+F\psi$.
 In fact,
general DG model describes six-parameter family of nonlinear
Hamiltonians, but the action of all its nonlinear terms on
realistic quantum systems is similar to $F$ of (\ref{BXB}), hence
for the start only this ansatz will be used in our calculations
\cite {DG2}. The choice of $\lambda$ of (\ref{BXB}) to be
imaginary was prompted by results of  nonrelativistic current
algebras \cite {DG}, but they doesn't have  mandatory character;
below we'll consider $F$ terms both with imaginary and real
$\lambda$.


Main properties of eq. (\ref{AXA}) for imaginary $\lambda$ were
studied in \cite {DG}, they can be promptly extended on
 real $\lambda$ and summarized for both cases as follows:
(a) The probability is conserved. (b) The equation is homogeneous.
(c) The equation is Euclidian- and time-translation invariant (
for $V=0$).
 (d) Noninteracting particle subsystem remain
uncorrelated (separation property). Distinct values of $\lambda$
can occur for different particle species.
 (e) Writing $<Q>=\int
\bar{\psi}\hat{Q}\psi d^3x$ for operator expectation value, in
particular,  since
 $ \int \bar{ \psi}F\psi d^3x=0 $, the energy functional for a
 solution of (\ref{AXA}) is $<i\hbar \partial_t>=<H_0>$, this property is quite 
important for our model.
  In this framework, it follows
  for $\vec{p}= -i\hbar\bigtriangledown $ and imaginary
 $\lambda$,
 $$
\frac{d}{dt}<\vec{r}>=\frac{<\vec{p}>}{m},
$$
 whereas for
 real  $\lambda$,
$$
\frac{ d}{dt}<\vec{r}>=\frac{1-2\lambda m}{m}<\vec{p}>
$$
For $V=0$, plane waves $\psi=\exp[i(\vec{k_0}\vec{r}-\omega t)]$
with $\omega=E\hbar$, $|\vec{k}_0|^2=2mE/\hbar$ are solutions
both for real and imaginary $\lambda$. For real $\lambda$, QM
continuity equation for probability density $\rho$  fulfilled,
but  density current acquires the form
$$
 \vec{j}=\frac{\hbar(1-2\lambda m)}{2mi}(\bar{\psi}\bigtriangledown\psi
 -\psi\bigtriangledown\bar{\psi })
$$
 For imaginary $\lambda$ the continuity equation  becomes of
Fokker-Plank type \cite {DG}.

As was mentioned above, the simple  quantum model of metastable
state decay describes it as  the particle tunneling  through the
potential barrier with suitable parameters \cite {Lan}. It's
natural to expect that for small $|\lambda|$  the tunneling
mechanism doesn't change principally, and  resulting 
difference  from standard QM solutions is small. Hence  the
nonlinear solutions can be treated as the perturbations to linear
solutions for the same system parameters. Exploiting this
auxiliary assumption, we'll find some unnormalised partial
solutions of nonlinear problem. Basing on them, we'll obtain the
exact solution of nonlinear problem for practically important
situations, which is independent of  mentioned auxiliary
assumption. To illustrate the influence of nonlinear DG term on
particle tunneling, consider 1-dimensional plane wave tunneling
 through the  potential barrier.
  Suppose that the rectangular barrier of the height $V_0$
  located between $x=0$ and $x=a$, and
the plane wave   with  energy $E < V_0$
 spreads from $x=-\infty$. Long-living metastable states
appear for small transmission coefficient $D \to 0$, which
corresponds to barrier width  $a\to \infty$ for fixed $E, V_0$.
For example, for $^{238}$U $\alpha$-decay $D\approx 10^{-37}$
\cite {Gam}. We'll study solutions of stationary  equation
(\ref{AXA}), basing on its asymptotic in this limit.
 Standard QM stationary solution for $x < 0$
$$
\psi_0(x)=\exp(ikx) + A \exp(-ikx)
$$
with $k=\frac{1}{\hbar}(2mE)^{\frac{1}{2}}$; for $a\to \infty$ it
gives $|A| \to 1$, i.e. it describes nearly complete wave
reflection from the barrier. Hence $\psi_0$ can be decomposed as
$\psi_0=\psi_{\infty}+\psi_d$ where the asymptotic state
$\psi_{\infty}={2}\cos(kx-\alpha_0)$, $\alpha_0=\arctan\chi_0/k$
where
 \begin {equation}
 \chi_0= \frac{1}{\hbar}{[2m(V_0-E)]}^{\frac{1}{2}} \label {AAZ}
\end {equation}
Then,  $\psi_d = A_d \exp(-ikx)$ where $|A_d| \simeq
\exp(-2\chi_0a)$, i.e. is exponentially small. In distinction, for
nonlinear equation
\begin {equation}
 E\psi=H_0\psi +F\psi   \label {XXZZ}
\end {equation}
the incoming and reflected waves suffer the  rescattering , hence
the stationary state $\psi \ne \psi_0$ for $x<0$. In our case, for
real $\lambda$,
  the stationary solution  can be obtained
 performing Ricatti
 transformation of solution of adjoined linear equation \cite
{DG,DG2}. Namely, for real solution $\eta(x)$ of such
Schroedinger equation the solution of corresponding nonlinear
equation
$$
            \psi=\eta \exp(\nu \ln \eta^2)
$$
where $\nu=\gamma(1-4\gamma)^{-1}$ and $\gamma=\lambda m$. Below
for brevity such exponential ansatz is replaced by corresponding
function rate. In particular, the asymptotic solution for $x\le
0$ and $a\to \infty$ can be written as
$$
    \psi_N=
    A_N[\cos(qx-\alpha)]^{\frac{1-2\gamma}{1-4\gamma}}
$$
where
 $$
q=\frac{1}{\hbar}\frac{[2mE(1-4\gamma)]^{\frac{1}{2}}}{1-2\gamma}.
$$
and
$$
   A_N=2^{\frac{1}{2}}{\pi}^{\frac{1}{4}}\Gamma^{-\frac{1}{2}}(\omega)
$$
where $\omega=\frac{2-6\gamma}{1-4\gamma}$; for small $|\gamma|$,
$\alpha \approx \alpha_0$. Plainly, $q \simeq k+o(\gamma^2)$,
hence below for small $|\gamma|$ they are taken to be equal.
 For imaginary $\lambda$ the corresponding nonlinear transformation
given in \cite {DG,DG2}, however, even for complete wave
reflection from the barrier the consistent asymptotic solution
 for $\psi_N$ doesn't exist, because $\psi_N$ phase singularities
 appear at its nodes.  In this case,
 the linear QM solution $\psi_{\infty}(x)$
 for the same system parameters can be used  as its approximation.
 For $x> a$, both for real and imaginary $\lambda$, the solution is
$$
\psi(x)=C^+ \exp(ikx) \simeq C^+\exp (iqx),
$$
To calculate the tunneling parameters, we need to use $\psi_N$
only for $x \to 0$, in this limit
$$
   \psi_N (x)\simeq
   A_N[1+\frac{\gamma}{1-4\gamma}\cos(qx-\alpha)]
$$
for finite but large $a$ the correction to it can be taken to be
equal to $\psi_d$, i.e. $\psi=\psi_N+\psi_d$. Next, it's necessary
to find the partial solution for $0 < x < a$ with $\psi(a) \to 0$
for $a \to \infty$, which should be the  main term of tunneling
state. For real $\lambda$, such solution of eq. (\ref {XXZZ}) is
$ \psi_1(x)=B_1\exp(-\chi x)$ where
\begin {equation}
    \chi
    =\frac{1}{\hbar}\frac{[2m(V_0-E)]^{\frac{1}{2}}}{(1-4\gamma)^{\frac{1}{2}}}
    \label {KKK}
\end {equation}

  Analogously to standard QM, another partial solution
describes the secondary term $\psi_2(x)=B_2\exp (\chi x)$, yet in
the linear case,
 $|B_2| \sim B_1\exp(-2\chi a)$, so $\psi_2$ is supposedly exponentially small in comparison with
$\psi_1$. Therefore, transmission coefficient can be estimated
with the good accuracy even accounting only main term $\psi_1$, it
gives $D_1=|B_1|^2\exp(-2\chi a)$, so that $D_1$  exponentially
depends on $\lambda$.

Due to  nonlinearity, the superpositions of two terms
$\psi_{1,2}$,
 in general, aren't solutions. Analytic solutions, which correspond
to such superpositions, exist in two cases only, defined by
$b_s=B_2/B_1$ ratio. First, for imaginary $b_s$ the solution is
just $\psi=\psi_1+ \psi_2$; second, for $b_s$ real
$$
    \psi(x)=B_1^{\frac{1-2\gamma}{1-4\gamma}}[
    \exp(-\chi_b
    x)+b_s\exp(\chi_b x)]
    ^{\frac{1-2\gamma}{1-4\gamma}}
$$
where
\begin {equation}
     \chi_b=\frac{1}{\hbar}\frac{[2m(V_0-E)(1-4\gamma]^{\frac{1}{2}}}{1-2\gamma}
     \label {XYZ}
\end {equation}
Meanwhile, for typical
 $\alpha$-decay parameters $E\approx V_0/2$, it corresponds to $\chi, q$ values such that $\chi \approx
 q$, hence we'll consider mainly such parameter range. It follows
 that  for $\chi=q$, there is the  solution of
 nonlinear problem with $\psi=\psi_1+\psi_2$ where
$$
      B_1=-\frac{2q(q-i\chi)}{(q^2+\chi^2)^2+(q-i\chi)^2\exp(-2\chi a)}
$$
and
 $$
B_2=B_1\frac{\chi^2-q^2+2iq\chi}{\chi^2+q^2}\exp(-2\chi a)
$$
so that $b_s$ is imaginary. $A_N,\,A_d,\,\alpha,\,$ and $C^+$ can
be calculated from $\psi$ continuity conditions for $0,a$. It's
notable that $B_{1,2}$ dependence on $\chi,q$ coincide with the
formulas for linear Hamiltonian \cite {Lan}. Denote
$$
   d_s=\frac{q^2-\chi^2}{q^2+\chi^2}
$$
Then, for $|d_s|<<1$ the obtained $\psi$ ansatz can be used as
first order approximation for nonlinear problem with $\chi\ne q$.
 If to denote the problem solution as $\psi_s$, then the next order term $\varphi^1$ for  $\psi_s=\psi+\varphi^1$  obeys to linear equation
\begin {equation}
(\frac{1}{2}-\gamma)\bigtriangledown^2\varphi^1-m(V_0-E+\lambda
\chi^2 )\varphi^1 +\frac{4\gamma \chi^2 d_s\exp(-2\chi a)
\psi}{\exp(-2\chi x) +\exp(2\chi x- 4\chi a)-2d_s\exp(-2\chi a)}=0
 \label {XYYZ}
\end {equation}
For  $\psi$ obtained above, this equation can be solved exactly.
Resulting $\psi_s$ can be extended further as
$\psi_s=\psi+\varphi^1+\varphi^2+...$ to higher $\gamma$ rate
$\varphi^2, \varphi^3,...$ terms, which also  would  obey to
linear equations.
In this case,  the resulting transmission coefficient $D_s\approx
2D_1$, so that $D_s$ also has the exponential dependence on
$\lambda$.

For imaginary $\lambda$ and small $\gamma$, the main term
$\psi_1=B_1\exp(-\chi_0\varpi x)$ where
$$
     \varpi= \frac{1+2m\lambda }{(1+4m^2|\lambda|^2)^{\frac{1}{2}}}
$$

 Transmission coefficient for main term is equal to
 $ D_1= |B_1|^2\exp(-2\chi_0 \upsilon a)$
 where
  $\upsilon= \rm{Re} \, \varpi$.
It supposes that $D_1$ dependence on $\lambda$ is less pronounced
than for real $\lambda$. Then, the secondary term
$\psi_2=B_2\exp(\chi_0\varpi x)$. Both for real and imaginary
$\lambda$, $C^+=\psi(a)\exp(-ika)$, which defines $\psi(x)$ for $x
> a$. Analogous considerations permit to derive next order
corrections $\varphi^2, \varphi^3,...$ to $\psi$ for imaginary
$\lambda$.


It's noteworthy  that considered nonlinear Hamiltonian term $F$
influences mainly the transitions between degenerate states, as
property (e) demonstrates.
 Due to it,  tunneling transmission coefficients
and related decay rates are  sensitive to the presence of
nonlinear
 terms in  evolution equation. Therefore, the  experimental studies of such
 process parameters
can be  important method of quantum nonlinearity search.

      \section {$\alpha$-decay oscillation model}

 Gamow theory of nucleus $\alpha$-decay
supposes that in the initial nuclei state,  free $\alpha$-particle
already exists inside the nucleus, but its total energy $E$ is
smaller than  maximal height of potential barrier $V(r)$ constituted by
nuclear forces and Coulomb  potential \cite {Gam}. Hence
$\alpha$-particle can leave nucleus volume only via quantum
tunneling through this barrier.
 For real nucleus,
The  barrier potential isn't rectangular, but has complicated form
described by some function $V(r)$ defined experimentally \cite
{Gam,New}. In this case, to calculate transmission rate in our
model,
 WKB approximation for Hamiltonian of eq. (\ref{AXA}) with nonlinear term of eq. (\ref{BXB})
 was used \cite{Lan}; its applicability to our nonlinear Hamiltonian
can be easily checked. The calculations described here  only for  real
$\lambda$, for imaginary $\lambda$ they are similar, it  supposed now that,
in principle, it can  depend on time, i.e. $\lambda=\lambda(t)$. 
 In this ansatz, 3-dimensional $\alpha$-particle wave function  reduced
to $\psi=\frac{1}{r}\exp(i\sigma/\hbar)$; function $\sigma(r)$
can be decomposed in $\hbar$ order $\sigma=\sigma_0-i\hbar\sigma_1+...$
\cite {Lan}.
Neglecting $\sigma_1,...$  and higher terms, the equation for main
term $\sigma_0$
\begin {equation}
       (\frac{1}{2m}-\lambda)[(\frac {\partial \sigma_0}{\partial {r}})^2 - i\hbar\frac {\partial \sigma_0^2}{\partial {r}^2} ]-
 \lambda|\frac {\partial \sigma_0}{\partial {r}}|^2 =E-V(r)  \label {AZZ2}
\end {equation}
Given $\alpha$-particle energy $E$, one can find the distances
$R_0, R_1$ from nucleus centre at which $V(R_{0,1})=E$. Then,
neglecting the term proportional to $\hbar$ this equation becomes
\begin {equation}
       (\frac{1}{2m}- \Lambda(r))   (\frac {\partial \sigma_0}{\partial {r}})^2=E-V(r)  \label {AZZ}
\end {equation}
where $\Lambda(r)=2\lambda$ for $R_0\leq r\leq R_1$, $\Lambda(r)=0$ for
$r<R_0$, $r>R_1$. Its solution for $R_0\leq r\leq R_1$ can be written as
$$
   \psi=\frac{1}{r}\exp(i\sigma_0/\hbar)=\frac{C_r}{r}\exp[-\frac{1}{\hbar}\int\limits_{R_0}^{r} p(\epsilon)
   d\epsilon]
$$
where $C_r$ is normalization constant,
$$
   p(\epsilon)=\frac{1}{\hbar}[\frac{2m(V(\epsilon)-E)}{1-4\gamma}]^{\frac{1}{2}}
$$
where $\gamma(t)=m\lambda(t)$. Account of higher order $\sigma$
terms doesn't change
 transmission coefficient which is equal to
\begin {equation}
  D=\exp[-\frac{2}{\hbar}\int\limits_{R_0}^{R_1}p(\epsilon)
   d\epsilon]=\exp[-\frac{\phi}{(1-4\gamma)^\frac{1}{2}}]    \label {ZXX}
\end {equation}
here $\phi$ is constant, whereas $\gamma$ can change in time. For
imaginary $\lambda$ the  calculations result in the same $D$
ansatz, but with
$$
 p(r)=\frac{1}{\hbar}[\frac{2m(V-E)}{1 +4m^2|\lambda|^2}]^{\frac{1}{2}}
$$
To calculate the nucleus life-time, $D$ should be multiplied by
the number of $\alpha$-particle kicks $n_d$ into  potential
barrier per second \cite {Gam}.

To study the decay parameter variations in external field, we'll
suppose now that  nonlinear Hamiltonian term $F$ depends on
external field. In our model, such field is gravity,
characterized  by its potential $ {U( \vec{R},t)}$. In this case,
$U$ should be accounted in evolution equation twice. First, $mU$
should be added to $H_0$, so that it changed to $H'_0=H_0+mU$;
second,
 nonlinear $H$ term $F$ can depend on $U$ or some its derivatives.
For  minimal modification of DG model we'll assume
 that for  $F$ ansatz of (\ref {BXB}) its
possible dependence on external field is restricted to parameter
$\lambda$ dependence: $\lambda=f(U)$, so now $\lambda$ isn't
constant, but the function of $\vec{R}$ and $t$. It supposed also
that $f \to 0$ for $U \to 0$, so that the free particle evolution is
linear.


Considered model doesn't permit to derive $\lambda$ dependence on
Sun gravity, but it can be obtained from its comparison with
experimental results for $^{214}$Po $\alpha$-decay \cite {Al}. We'll suppose
that $\lambda$ is function of potential $U(\vec{R}_n,t)$
 where $\vec
{R}_n$ is nucleus coordinate in Sun reference frame (SRF).
 As
follows from eq. (\ref{ZXX}) for small $\lambda$
$$
       D \approx (1+2\phi\gamma) \exp(-\phi)
$$
For $^{214}$Po decay, its life-time $\tau_0=16.4*10^{-6}$ $sec$,
model estimate gives $\phi \approx 60$. For annual $\tau$
variation the best fit for 3 year exposition has main harmonics
$$
               \tau_a(t)=\tau_0 [1+A_a\sin w_a(t+\varphi_a)]
$$
where $t$ defined in days, $A_a=9.8*10^{-4}$, $w_a=2\pi/365$,
$\varphi_a=174$  days \cite{Al}. Remind that Earth orbit is
elliptic,
 the minimal distance from Sun is at about January 3 and maximal at
about July 5, maximal/minimal orbit radius difference is about
$3*10^{-2}$ \cite {Wei}. Plainly, the minima and maxima of $U$
time derivative $\partial_t U$ will be located approximately in
the middle between these dates, i.e. about April 5 and October 3,
correspondingly. In general, this dependence described as
$$
               \partial_t U = K^a \sin w_a(t+\varphi_u)
$$
  here $\varphi_u=185$ days, $K^a=1.5\, m^2/sec$, as the result,
such model $\varphi_u$ value in a good agreement with experimental
$\varphi_a$ value. Thereon, it means that the plausible data fit
is $\lambda(t)=g\partial_t U$, where $g$ is interaction constant,
which can be found from the data for $^{214}$Po decay. It follows
from the assumed equality of oscillation amplitudes $A_a=2\phi
mgK^a$ that the resulting $g=.35*10^{-8}$ $\frac{sec^3}{m^2 MeV}$.

Another experimentally found harmonic corresponds to daily variations
with best fit
$$
    \tau_d(t)=\tau_0 [1+A_d\sin w_d(t+\varphi_d)]
$$
where $t$ defined in hours, $A_d=8.3*10^{-4}$, $w_d=2\pi/24$,
$\varphi_d=12$  hours \cite{Al}. Such oscillation can be
attributed to variation of Sun gravity due to daily lab. rotation
 around Earth axe. It's easy to check that nucleus life-time
 dependence also coincides with  $\partial_t U$ time dependence with
 high precision. Really, it   described as
 $$
     \partial_t U = K^e \sin w_d(t+\varphi_e)
 $$
with $\varphi_e=12$ hours, $K^e=.9 \,m^2/sec$ \cite {Wei}. It
follows that $A_d=2\phi mgK^e$; if to substitute in this equality
$g$ value, calculated above, it gives $A_d=5.5*10^{-4}$, which is
in a reasonable agreement with its experimental value. It's
possible also that $\tau_{a,d}$  can depend on some other orbit
parameter or $U$ derivative, in particular, on lab. velocity in
SRF or some absolute reference frame \cite {Al}; such options
will be considered elsewhere.

\section{Nonlinearity, Nonlocality and Causality  }
\label{sec-1}

In this paper, we studied hypothetical  nonlinear corrections to
standard QM description of system interaction with external
fields. It's notable that in nonrelativistic QM, the ansatz for
system interaction with massless fields, such as gravity and
electromagnetism, is stipulated, in fact, by Bohr
quantum-to-classical correspondence principle \cite {Jauch,Fay}.
However, in the last sixty years many new physical concepts were
discovered, which presumably are independent of it. The
illustrative example is, in our opinion, quantum chromodynamics,
the theory derived just from experimental facts with no reference
to correspondence principle \cite {Li}. In the same vein, there
is no obvious prohibition on the existence of additional
Hamiltonian terms for the system interaction with massless
fields. Such terms can have strictly quantum origin and disappear
in classical limit, their existence should be verified in
dedicated experiments. To study their general features, we
considered the simple nonrelativistic model, which includes the
additional terms  for the interaction of quantum systems with
gravitational field. Account of these terms permits to describe
with the  good accuracy the annual and daily oscillations of
$^{214}$Po $\alpha$-decay parameters observed in the experiment.


 It was argued earlier
that  QM nonlinearity violates relativistic causality for
multiparticle systems, in particular, it permits the superluminal
signaling for EPR-Bohm pair states \cite {Gis,Svet}. However,
this conclusion was objected and still disputed \cite {Jor}.
Plainly, these arguments would be even more controversial, if
nonlinear effects exist only inside the  field volume.  In
particular, the metastable state in external field can be
considered as the open system, yet it was shown that the
superluminal signaling between such systems is impossible \cite
{Cz}. Moreover, heavy nucleus is strongly-bound system, so it's
unclear whether it's possible to prepare  the entangled state of
two $\alpha$-particles located initially inside two different
nuclei.
 It was shown that in our
model the standard relation between average system momentum and
velocity can be violated and differ from particle mass. However,
because in our model this ratio depends on external gravitational
field,  then for Sun gravity influence it can depend on time of
day or year season, hence its tests demand the special subtle
experiments.

It was proposed by many authors  that  gravity is emergent
(induced) theory and originates, in fact, from  some nonlocal
field theory (see \cite {Diaz} and refs therein). In this
framework, it was supposed that gravity  can be effectively
described by multilocal (collective) field $\Phi_n(x_1,...,x_n)$
or the array of such field modes $\{\Phi_1,...,\Phi_n\}$. It was
shown that bilocal scalar field $\Phi_2$ reproduces the classical
gravity effects up to the second order \cite {Diaz}. Such bilocal
field $\Phi_2$ presumably can interact with bilocal operators of
massive fields, in particular, such field  can be the
nonrelativistic particle system. Such interaction doesn't
violate causality, if for the pair of separated quantum objects it
 influences only their bilocal observables of EPR-Bohm type \cite
{Jauch,Gis}. The simple example of such observable is the spin projection difference for
two fermions.

In our phenomenological model, it assumed that in infrared limit
the gravity effectively described by two terms $\{ \Phi_1, \Phi_2
\}$ where $\Phi_1=U(\vec{r})$ is  standard Newtonian potential.
Denote as $\vec{r}_1$ the coordinate of $\alpha$-particle,
$\vec{r}_2$ the coordinate of remnant nucleus centre of mass, and
 $\vec{r}_s=\vec{r}_1-\vec{r}_2$.
For considered $\alpha$-decay model, the joint state of remnant
nucleus and $\alpha$-particle is entangled, their bilocal
observable $\vec {r}_s$ is of EPR-Bohm type.
It's notable that it's equivalent to the basic   coordinates of
bilocal field which described  as $\Phi_2(\vec{R}_a, \vec{r}_s)$
where $\vec{R}_a=\frac{1}{2}(\vec{r}_1+\vec{r}_2)$ \cite {Diaz}.
If  gravity field is local then  for $<|\vec{r}_s|><<
|\vec{R}_n|$ it will act mainly on nucleus total state
$\Psi(\vec{R}_n)$, its influence on on nucleus internal state
$\psi(\vec{r}_s)$ will be negligible. Only bilocal field can
change it, and as follows from our analysis of $^{214}$Po
$\alpha$-decay data, it's plausible that for bilocal scalar field
 $\Phi_2 \sim \partial_tU$. As was argued above, the nonlinear
 term should act on $\vec {r}_s$, hence assuming that $\Phi_2$  factorized from
 such
 operator, it follows
$$
  {F} \sim \partial_tU \, G(\frac{\partial}{\partial
    \vec{r}_s})
$$
where G is some operator function. In this case, the analogue of
D-G ansatz will contain
  variable mutiplier $k_b\partial_tU$ in place of fixed $\lambda$, where $k_b$ is arbitrary constant.
  Then, in distinction from initial nonlinear term of eq. (\ref{BXB}), the corresponding nonlinear term of our Hamiltonian for nucleus
  $\alpha$-decay becomes
$$
    F=k_b\partial_tU(\vec{R_n})(\frac{\partial^2}{\partial
    \vec{r}_s^2}+\frac{1}{|\psi|^2}|\frac{\partial \psi}{\partial
    \vec{r}_s}|^2)
$$
where $k_b$ is an arbitrary constant.
  It means that $\Phi_2$ interaction with the nucleus
described by nonlinear operator;  as the result, $\alpha$-particle
transmission coefficient $D(t)$ can oscillate around its constant
value defined by Gamow theory. For large $|\vec{r}_s|$ it can be
supposed that $\partial_tU(\vec{R_n})$ should be replaced by
\begin {equation}
    \Phi_2=\frac{1}{2^\frac{1}{3}}\{\partial_tU(\vec{r}_1)\partial_tU(\vec{r}_2)
    [\partial_tU(\vec{r}_1)+\partial_tU(\vec{r}_2)]\}^\frac{1}{3}
    \label{ZZZZ}
\end {equation}
As follows from equivalence principle, in lab.
 reference frame, located on Earth surface, Sun gravitation
 potential $U'(\vec{R}_c)\approx 0$, yet $\partial_{\vec{r}}U' \ne
 0$. It means that nuclear decay process  violates  equivalence
 principle, however, some theories of emergent (induced)
 gravity predict that it can be violated in  quantum processes
 at small scale \cite {Diaz,Raam}.
  In addition, other results for $^{214}$Po
 $\alpha$-decay seems to support such conclusion. Namely, beside
 described life-time oscillations, these data contains also
 harmonics with period $24$ hours $50$ minutes, which is equal to
 lunar day duration and so can be related to well-known moon gravity effects
\cite {Al}.
Studies in quantum gravity supposes that this theory can be
similar to QFT with massless messenger called graviton (\cite
{Diaz,Raam} and refs. therein). Notorious example of massless
messenger formalism gives QED,  it's well known yet that in its
nonrelativistic limit there are some electromagnetic effects,
like Casimir effect or Lamb shift, which can't be described by
Scroedinger equation, but only via accounting higher order QED
terms \cite {Li}. It seems possible that the observed decay
oscillations can have analogous origin corresponding to  the
nonzero infrared limit of some hypothetical quantum gravity
terms, which can appear, in particular, if gravity is
nonperturbative theory. 

 Considered QM nonlinearity supposedly has universal
character, so beside nucleus decays, such temporary variations
under influence of Sun gravity can be observed, in principle, for
other systems in which metastable states and tunneling play
important role. In particular, it can be some chemical reactions,
molecular absorption by solids and liquids, etc.
It's worth to
notice specially its possible role in some biological processes.
 Multiple publications indicate that some cosmophysical effects
influence also biological system development and functioning (see \cite {Hay,Kol} and refs. therein).
Of them, it's worth to notice specially the observed influence of  moon tide gravity variations $\delta g$ on  seedling bioluminescence rate and tree stem diameter variations \cite {Bar,Mor}. There is no consistent explanation up to now how such small gravitational force variations $\frac{\delta g}{g}\sim 10^{-7}$ can seriously affect such subtle biological processes. It's worth to notice specially that bioluminescence  data show the essential intensity dependence on  $\delta g$ time derivative  \cite {Mor,Gal2}. Meanwhile, as was shown above,  our  model of  gravitational field interaction with quantum systems predicts  similar gravity influence on arbitrary quantum systems proportional to $U$ time derivatives.
It's  established now that such bioluminescence stipulated mainly  by biochemical reactions of protein oxidation \cite {Mor,Gal}. Hence it can be assumed that observed $\delta g$ time derivative dependence owed to such nonlinear gravity interaction with molecular states involved in these reactions.






\end {document}